\documentclass[aps,showpacs,showkeys,superscriptaddress,nofootinbib,preprint,11pt]{revtex4}
\usepackage{graphicx}
\usepackage{float}
\begin{document}

\title{Determination of $sin^{2}\theta_W$ using $\nu(\bar\nu)$-Nucleus scattering}

\author{H. Haider}
\affiliation{Department of Physics, Aligarh Muslim University, Aligarh-202 002, India}

\author{I. Ruiz Simo}
\affiliation{Dipartimento di Fisica, Universit\'a degli studi di Trento
Via Sommarive 14, Povo (Trento)
I-38123, Italy}

\author{M. Sajjad Athar}
\affiliation{Department of Physics, Aligarh Muslim University, Aligarh-202 002, India}

\begin{abstract}
We have studied nonisoscalarity and medium effects in the extraction of weak mixing angle using Paschos and Wolfenstein relation in the iron nucleus.
Paschos and Wolfenstein(PW) relation is valid for an isoscalar target. We have modified the PW relation for nonisoscalar target as well as incorporated the 
medium effects like Pauli blocking, Fermi motion, nuclear binding energy and pion rho cloud contributions. In our calculations we have used the relativistic nuclear spectral function which includes nucleon correlations. 
Finally local density approximation is applied to translate the numerical results to the finite nuclei. We have studied the dependence of $sin^{2}\theta_W$
on Bjorken variables $x$ and $y$, four momentum transfer square ($Q^2$), energy of the neutrino and antineutrino, and effect of excess neutrons over protons in the nuclear target.
\end{abstract}
\maketitle


\section{Introduction}
NuTeV Collaboration has obtained $sin^{2}\theta_W$ using iron nuclear target and found $\sin^2 \theta_W$ to be $0.2277\pm0.0004$, which is
 3 standard deviations above the global fit of $sin^2 \theta_W=0.2227\pm0.0004$ and this is known as NuTeV anomaly~\cite{Zeller}.
To resolve this discrepancy, explanations within and outside the standard model of electroweak interactions have been looked for.
In this paper we have studied the impact of nuclear effects and nonisoscalarity corrections on the extraction of the weak-mixing angle $sin^{2}\theta_W$, using
antineutrino and neutrino nucleus scattering. The calculations have been performed in a theoretical model using relativistic nuclear spectral functions which incorporate Pauli blocking, Fermi motion,
binding energy and nucleon correlations. We have also included the pion and rho meson cloud contributions calculated from a microscopic model for meson-nucleus
self-energies. The details of the model are given in Refs. \cite{sajjadnpa29}-\cite{prc}.\

Paschos and Wolfenstein(PW) demonstrated that for an isoscalar target the ratio of neutral current to charged current cross sections is given by~\cite{prc}:
\begin{eqnarray} \label{pwrelation}
R_{PW}=\frac{\sigma(\nu_\mu~N \rightarrow \nu_\mu~X)~-~\sigma(\bar\nu_\mu~N \rightarrow \bar\nu_\mu~X)}{\sigma(\nu_\mu~N \rightarrow \mu^-~X)~-~\sigma(\bar\nu_\mu~N \rightarrow \mu^+~X)}=\frac{1}{2}~-~\sin^2 \theta_W
\end{eqnarray}
The above relation assumes no contribution from heavy quarks, symmetry between the strange quark and anti-strange quark and considers no medium effect and no contribution from outside the standard model.
NuTeV Collaboration~\cite{Zeller} has extracted the weak mixing angle using the differential scattering cross section in the PW relation as Eq. (\ref{pwrelation}) is also valid for 
differential cross section for an isoscalar nuclear target.
\section{Formalism}
The differential cross section for charged current (anti)neutrino interaction with a nucleus is written as \cite{prc84}:
\begin{footnotesize} 
\begin{eqnarray}
\frac{d^2\sigma^{\nu(\bar{\nu})A}_{CC}}{dE'\;d\Omega'}=\frac{G^2_F}{(2\pi)^2} \; \frac{\left|\vec{k}'\right|}{\left|\vec{k}\right|} \; \left(\frac{m^2_W}{q^2-m^2_W}\right)^2 \;
 L^{\alpha\beta}_{\nu(\bar{\nu})} \; W^{\nu(\bar{\nu})A}_{\alpha\beta} 
\end{eqnarray}
\end{footnotesize}
where $G_F$ is the Fermi coupling constant, $m_W$ is the mass of the W boson and $l(=e,\,\mu)$ is a lepton. 
$k$ and $k^\prime$ are the four momentum of the incoming neutrino and outgoing lepton respectively and $q=k-k^\prime$ is the four momentum transfer.

$L^{\alpha\beta}_{\nu(\bar{\nu})}$ is the leptonic tensor tensor which is given as
\begin{eqnarray}
L^{\alpha\beta}_{\nu(\bar{\nu})}(k,k')&=&k^\alpha k'^\beta+k'^\alpha k^\beta-g^{\alpha\beta}(k\cdot k')\pm i\;\epsilon^{\alpha\beta\mu\nu}k_\mu k'_\nu \label{eq:leptonic_tensor}\\ \nonumber
\end{eqnarray}
where plus sign is for antineutrino and minus sign is for neutrino.

In the local-density approximation the nuclear hadronic tensor $W^{\nu(\bar{\nu})A}_{\alpha\beta}$ can be  written as a convolution of the nucleonic
hadronic tensor with the hole spectral function. For a N$\neq$Z nucleus like $^{56}Fe$, we have considered separate distributions of Fermi sea
for protons and neutrons\cite{prc85}:\
\begin{eqnarray}
W^{\nu(\bar{\nu})A}_{\alpha\beta}&=&2\left\langle \int^{\mu_p}_{-\infty}dp^0\; S^{p}_{h}(p^0,\mathbf{p},k_{F,p}) ~ W^{\nu(\bar{\nu})p}_{\alpha\beta} \right\rangle +
 \left\langle 2 \int^{\mu_n}_{-\infty}dp^0\; S^{n}_{h}(p^0,\mathbf{p},k_{F,n})~ W^{\nu(\bar{\nu})n}_{\alpha\beta} \right\rangle ,
\end{eqnarray}
 $S_h^p$ and $S_h^n$ are the two different spectral functions, each one of them normalized to the number of protons or neutrons in the nuclear target
 and are functions of Fermi momentum of protons
and neutrons respectively which are given by $k_{F,p}=\frac{3\pi^2\rho_p}{3}$ and $k_{F,n}=\frac{3\pi^2\rho_n}{3}$ and
\[\left\langle  \;\;  \right\rangle=\int d^{3}r\int\frac{d^{3}p}{(2\pi)^{3}}\frac{M}{E(\vec{p})}.......\]
PW ratio for nonisoscalar nucleus($R^{PW}_{NI}$) in our model may be written in terms of PW ratio for  the isoscalar nucleus($R^{PW}_{I}$) and
 the correction due to nonisoscalarity($\delta R$)~\cite{prc}:    
\begin{eqnarray}\label{delta_modified}
 R^{PW}_{NI}&=&R^{PW}_{I}\;+\;\delta R=\frac{1}{2}-\sin^{2}\theta_W + \delta R
\end{eqnarray}
where
\begin{eqnarray}
\delta R=\delta R_{1}\;+\;\delta R_{2}\;+\;\delta R_{3} \label{deltaRNI}
\end{eqnarray}
with
\begin{footnotesize} 
\begin{eqnarray}\label{deltar1}
  \delta R_{1}=\frac{1}{D}\frac{1}{3}~sin^2\theta_w \times \left\langle\frac{\delta}{2V}\frac{\pi^{2}}{k^{2}_F}\int_{-\infty}^{\mu}dp^{0}\,
\frac{\partial S_h(p^{0},\vec{p},k)}{\partial k}\Big|_{k=k_F}\frac{2}{\gamma} \frac{p_0 \gamma - p_z}{p_0 - p_z \gamma}(u_v-d_v)\right\rangle,
\end{eqnarray}

 \begin{eqnarray}\label{deltar2}
\delta R_{2}=-\left(\frac{1}{2}-sin^2\theta_W \right)\,
\frac{y}{2-y} \frac{1}{D}\;\left\langle \frac{\delta}{2V}\frac{\pi^{2}}{k^{2}_F}\int_{-\infty}^{\mu}dp^{0}\,
\frac{\partial S_h(p^{0},\vec{p},k)}{\partial k}\Big|_{k=k_F}~\left[2+\frac{p_x^{2}}{(p\cdot q)} 4x_N\right]~(u_v-d_v)\right\rangle,
\end{eqnarray}

\begin{eqnarray}\label{deltar3}
 \delta R_{3}=-\left(\frac{1}{2}-sin^2\theta_W \right)\,\,
\frac{1-y-\frac{Mxy}{2E_\nu}}{xy\left(1-\frac{y}{2}\right)}
\frac{1}{D}\;\left\langle \frac{\delta}{2V}\frac{\pi^{2}}{k^{2}_F}\int_{-\infty}^{\mu}dp^{0}\,
\frac{\partial S_h(p^{0},\vec{p},k)}{\partial k}\Big|_{k=k_F}~G~x_N~(u_v-d_v)\right\rangle, \nonumber\\
\end{eqnarray}
\end{footnotesize}
\begin{footnotesize}
\begin{eqnarray*}
 G(p^{0},\vec{p})\equiv\frac{q^{0}\left[ q^{2}(\vec{p}\,^{2}+2(p^{0})^{2}-p_z^{2})-
2(q^{0})^{2}\left( (p^{0})^{2}+p_z^{2}\right)+4p^{0}q^{0}p_z\sqrt{(q^{0})^{2}-q^{2}}\right] }
{2M\left(q^{2}-(q^{0})^{2}\right)\cdot(p\cdot q)},
\end{eqnarray*}

\[D=\left\langle
\int_{-\infty}^{\mu}dp^{0}\,S_h(p^{0},\vec{p},k)\,\frac{2}{\gamma} \frac{p_0 \gamma - p_z}{p_0 - p_z \gamma}
\left(u_v+d_v \right)\right\rangle,\]

\begin{eqnarray*}	
\gamma=\frac{q_z}{q^0}=\left(1+\frac{4M^2x^2}{Q^2}\right)^{1/2}\,,~~x_N=\frac{Q^2}{2(p^0q^0-p_zq_z)}\,,~~\frac{\delta}{V}=\frac{N-Z}{V}=\rho_n(r)-\rho_p(r)~~\rm{and}
\end{eqnarray*}
\end{footnotesize}

$u_v$ and $d_v$ are the up and down valence quark distributions respectively. $E_\nu$ is the neutrino energy, M is the mass of nucleon.
We have used Eqs. (\ref{deltar1}), (\ref{deltar2}), and (\ref{deltar3}) in Eq. (\ref{deltaRNI}) to obtain the nonisoscalar correction to the PW relation and then calculated
$sin^{2}\theta_W$ using Eq.(\ref{delta_modified}).
\begin{figure}[H]
  \includegraphics[height=.25\textheight,width=13cm]{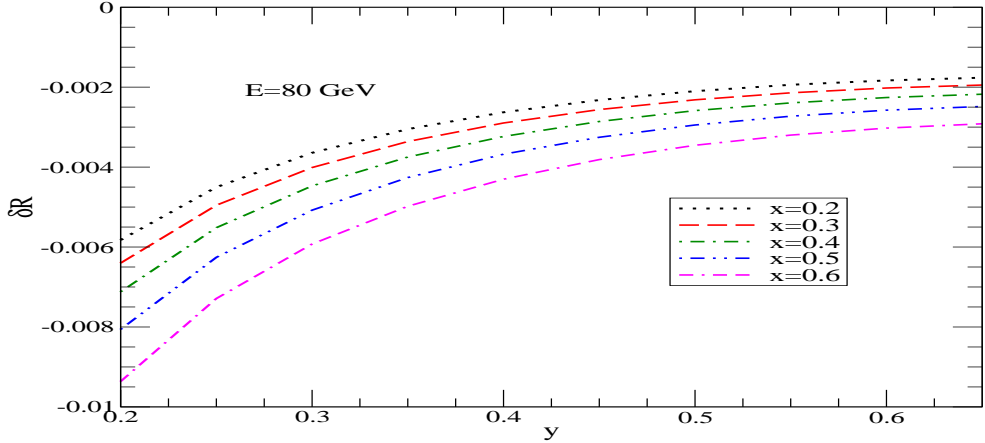}
  \caption{Nonisoscalar correction($\delta R$) vs y at different values of $x$ for $E = 80$ GeV.}
\label{fig1}
\end{figure} 
\begin{figure}[H]
\includegraphics[height=.25\textheight,,width=13cm]{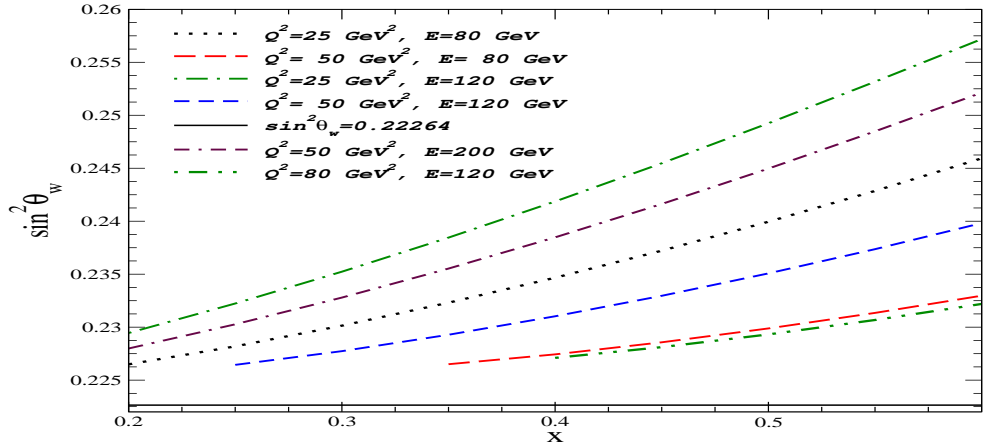}
\caption{$sin^{2}\theta_w$ vs $x$  in $^{56}$Fe treating it to be nonisoscalar target.
The results are shown at different values of $Q^2$ and E.}
\label{fig3}
\end{figure} 

\section{Results and Discussion}
To see the effect of nonisoscalarity in iron target, we have plotted $\delta R$ vs $y$ for  different values of $x$ at  anti(neutrino) energy $E= 80$ GeV in Fig. \ref{fig1}.
We find that the effect of non-isoscalarity is large at low $y$ and high $x$ which decreases with the increase in the value of $y$.
This effect is smaller at low values of $x$. In Fig. \ref{fig3} we have presented the dependence of isoscalar corrected $sin^{2}\theta_W$ on $E$ and $Q^2$
as well as the effect of medium on the weak mixing angle. We observe that at low $x$ for  $E=80$ GeV and $Q^2$=25 $GeV^2$,
$sin^{2}\theta_W$ is almost close to the standard value.  The value of  $sin^{2}\theta_W$ changes significantly with the change in $E$ , $Q^2$ and  $x$.
Furthermore, we observe that there is a strong dependence on nuclear medium effects as well as nonisoscalarity corrections in the different regions of $x$ and $Q^2$.
Therefore, nonisoscalarity and nuclear medium effects are important in the extraction of $sin^{2}\theta_W$, as well as it depends on at what (anti)neutrino
energies the evaluation is performed for a fixed $x$ and $Q^2$.





\end{document}